\documentstyle[aps,pra,12pt,eqsecnum]{revtex}
\begin{document}
\hfill
solv-int/9508004\\[10pt]
\title{
Numerical Studies of Localized Structures on an Uneven Bottom
in Two Dimensions}
\author{Tetsu YAJIMA
\footnote{E-mail: yajimat@mmm.t.u-tokyo.ac.jp }}
\address{Department of Applied Physics, Faculty of Engineering,
                      University of Tokyo,\\
         Hongo 7-3-1, Bunkyo-ku, Tokyo 113, JAPAN}
\author{Katsuhiro NISHINARI
\footnote{E-mail: nishi@dips.dgw.yz.yamagata-u.ac.jp}}
\address{Faculty of Engineering, Yamagata University,\\
         Jonan 4-3-16, Yonezawa, Yamagata 992, JAPAN}
\maketitle
\bigskip\bigskip
\vfill\vfill
\begin{abstract}
The Davey-Stewartson (DS) equations with a perturbation term
are presented by taking a fluid system as an example
on an uneven bottom.
Stability of dromions, solutions of the DS equations with
localized structures, against the perturbation is investigated
numerically.
Dromions decay exponentially under an effect of the perturbation,
while they travel stably after the effect disappears.
The decay ratio of dromions is found to have relation to
velocities of dromions.
The important role played by the mean flow,
which acts as an external force to the system, is discussed.
These results show that dromions are quite stable as a
localized structure in two dimensions,
and they are expected to observed in various physical systems
such as fluid or plasma systems.
\end{abstract}
\medskip\noindent
\hrule
\newpage
\section{Introduction}
Nonlinear problems in one spatial dimension have been studied
intensively and there are a lot of fruitful results from these
researches.
One of them is the study on nonlinear integrable equations,
such as Korteweg - de Vries equation or nonlinear
Schr\"odinger equation,
and soliton is derived as a solution of those equations.
Solitons are localized structures with exponentially
decaying bases, stable under interactions between each other,
and have properties both of wave and of particles.
Many problems have been discussed for these solutions from the
viewpoints of stability under many kinds of
perturbations\cite{MM,Kaku,AO,KM,IW}.
\par
On the other hand, in the systems in two spatial dimensions
or more, few equations are known to have localized solutions.
The Kadomtsev-Petviashvili (KP) equation is one of the
equations in two dimensions which have soliton solutions,
but these solitons localize only in one direction.
None of the known solutions of the KP equation other than
the soliton solutions, such as lump solution or the resonance
solution\cite{SA}, 
localize in two dimensions with exponentially decaying base
either.
The solutions of the KP equation cannot be recognized as
localized structures in two dimensions.
\par
Apart from the usual soliton solutions, we can nominate
dromion\cite{BLMP,FS1,FS2,HH},
which is a exact solution recently found in the
Davey-Stweartson 1 (DS1) equations\cite{DS,AS}:
\begin{equation}
\left\{\matrix{
iA_t+A_{xx}+A_{yy}+(U+V)A=0, \cr
U_y=(|A|^2)_x,\quad V_x=(|A|^2)_y, \hfill\cr}\right.
\label{DS1-ALL}
\end{equation}
as a localized structure in two dimensions.
For this solution, the envelope of $|A|$ has a localized
structure in two dimensions with a base which decays
exponentially.
\par
Dromions have quite different properties compared to solitons,
although both of them are localized spatially.
For example, if two dromions have interaction between each
other, the number of pulses generally may not be conserved.
Hence we cannot infer behaviors of dromions from the analogies
of those of solitons.
Moreover, detailed discussions on stability against
perturbations of dromions are quite few compared to those on
solitons in one dimensional space.
For these reasons, numerical analyses on the behaviors of
dromions against various perturbations are quite necessary.
\par
In this paper, our purpose is to take perturbations to the
DS1 equations into account
and clarify how they affect on the dromion solutions.
We consider perturbation terms due to inhomogeneity of
the bottom of the fluid in two dimensions.
This paper is organized as follows.
In the next section, we derive the DS equations with a
perturbation term and consider the effect on dromions.
In \S 3, detailed studies on the stability in typical
situations is presented by numerical analyses.
Discussions and concluding remarks are presented in the
final section.

\section{Model Equations}
In the previous paper\cite{YN}, we presented the perturbed
DS1 equations by using the reductive perturbation method.
First we briefly summarize the derivation
in order to see the meaning of the perturbation term,
and then discuss the effect of it.
\par
We consider a system of irrotational incompressible fluid
on an uneven bottom in three dimensions.
The velocity potential will be expressed as $\phi(x,y,z,t)$.
Then the basic equations are the Laplace equation
\begin{mathletters}
\label{BASEQ_all}
\begin{equation}
\Delta\phi=0, \qquad (-H<z<\zeta),
\label{BASEQ_laplace}
\end{equation}
and the boundary conditions at the surface and the bottom of
the fluid.
These are
\begin{eqnarray}
&&H_x\phi_x+H_y\phi_y+\phi_z=0, \quad (z=-H),
\label{BASEQ_bottom}\\
&&\phi_z-\zeta_t-\phi_x\zeta_x-\phi_y\zeta_y=0, \quad (z=\zeta),
\label{BASEQ_surface}\\
&&\phi_t+g\zeta+(\nabla\phi)^2/2=T/R\rho, \quad (z=\zeta),
\label{BASEQ_bernoulli}
\end{eqnarray}
\end{mathletters}%
where we have expressed the density of fluid as $\rho$,
the bottom as $z=-H(x,y,t)$ and the surface as$z=\zeta(x,y,t)$.
The quantity $g$ is the acceleration due to the gravity,
$T$ the surface tension and $R$ the curvature of the surface,
which is given by
$$
R={\displaystyle\strut
   (1+\zeta_x^2+\zeta_y^2)^{3/2}\over
   (1+\zeta_x^2)\zeta_{yy}+(1+\zeta_y^2)\zeta_{xx}
     -2\zeta_x\zeta_y\zeta_{xy}}.
$$
Now we are going to derive nonlinear amplitude evolution
equations for the velocity potential by means of the reductive
perturbation method\cite{DS,TW,TY}.
The quantity $H$ is not a constant, but we consider that its
variation is slow and small in space.
We also assume the spatial variance of $\zeta$ is sufficiently
slow compared with the characteristic length of the
localized structures.
Then let us consider the deviation of $(\phi,\zeta,H)$
around the stationary values $(0,0,H^{(0)})$,
where $H^{(0)}$ is a constant.
We expand them as
\begin{mathletters}
\label{EXPAND_all}
\begin{eqnarray}
&&\phi=\sum_{n=1}^\infty\sum_{\ell=-n}^n
       \phi_\ell^{(n)}(\xi,\eta,z,\tau)
       \varepsilon^ne^{i(kx-\omega t)\ell},
\label{EXPAND_phi}\\
&&\zeta=\sum_{n=1}^\infty\sum_{\ell=-n}^n
       \zeta_\ell^{(n)}(\xi,\eta,z,\tau)
       \varepsilon^ne^{i(kx-\omega t)\ell},
\label{EXPAND_zeta}\\
&&H=H^{(0)}+\sum_{n=1}^\infty
       H^{(n)}(\xi,\eta,\tau)\varepsilon^n.
\label{EXPAND_bot}
\end{eqnarray}
\end{mathletters}%
We have introduced expand parameter $\varepsilon$ and
new independent variables $\xi=\varepsilon(x-vt)$,
$\eta=\varepsilon y$, $z=z$ and $\tau=\varepsilon^2 t$,
where $v$ is the group velocity.
We assume relations
$\phi^{(n)}_{-\ell}=\phi^{(n)\ast}_{\ell}$ and
$\zeta^{(n)}_{-\ell}=\zeta^{(n)\ast}_{\ell}$
from the reality of the physical quantities.
The asterisk indicates complex conjugate.
Substituting these into (\ref{BASEQ_all})
and writing $K\equiv k^2T/\rho g$ and
$\sigma\equiv\tanh kH^{(0)}$,
we can get the dispersion equation
$$
\omega^2=gk\sigma(1+K),
$$
and relations between expansion amplitudes $\phi^{(n)}_{\ell}$
and $\zeta^{(n)}_{\ell}$ in (\ref{EXPAND_phi}) and
(\ref{EXPAND_zeta}).
The main results up to $O(\varepsilon^2)$ are listed in the
Table \ref{TAB-RES-RPM}.
In the next order, we have a set of equations whose dependent
variables are $\phi^{(1)}_0$ and $\phi^{(1)}_1$.
we introduce new dependent variables $A$ and $Q$ instead of
$\phi_\ell^{(n)}$:
\begin{eqnarray*}
A&=&A_1, \\
Q&=&v\{A_{0\xi}[gH^{(0)}-v^2](1+K)^2/k^2
       +|A_1|^2[v(1+\sigma^2)(1+K)^2+(1+K)
           -2K\omega/k]\},
\end{eqnarray*}
where $A_0$ and $A_1$ are listed in the Table \ref{TAB-RES-RPM}.
Applying suitable scaling and 45$^\circ$ rotation of the
coordinates, we have
\begin{mathletters}
\label{all_DSP0}
\begin{eqnarray}
&&i A_t+A_{\xi\xi}+A_{\eta\eta}-2 |A|^2A
   + A(Q_\xi+Q_\eta) = - iJ(\xi,\eta)A,
\label{OUR_EQ1}\\
&&Q_{\xi\eta}=(|A|^2)_{\xi}+(|A|^2)_{\eta},
\label{OUR_EQ2}
\end{eqnarray}
\end{mathletters}%
when the coefficients of the amplitudes satisfies a certain
relation.
The function $J(\xi,\eta)$ is real and proportional to
$H^{(1)}_\xi$, with positive proportional coefficient.
These are the DS1 equations with an extra term in the
right-hand side of (\ref{OUR_EQ1}),
which comes from the inhomogeneity of the system.
\par
As we have discussed in the Ref.\ \CITE{YN},
we take $H^{(1)}$ proportional to $\tanh(p\xi+q\eta)$
($p$ and $q$ are real constants).
This means there is a mild step into one direction on the bottom.
In this case, a simple calculation shows that the perturbation
term becomes
$$
iJ(\xi,\eta)A=i\alpha{\rm sech}^2(p\xi+q\eta).
$$
The quantity $\alpha$ indicates the degree of inhomogeneity,
and $p$ and $q$ determines the width and the direction of
the step.
\par
We consider a traveling wave solution, and set
$A\sim F(t)e^{i(k_x\xi+k_y\eta-\omega t)}$.
This wave travels away from the origin
in the first quadrant asymptotically.
Since we have assumed that $H$ varies slowly in space,
we can neglect higher order spatial derivatives of $H^{(1)}$.
Ignoring the nonlinear term in (\ref{OUR_EQ1}),
we have $\omega=(k_x^2+k_y^2)$ and
$F=\exp[-\alpha J(\xi,\eta) t]$\cite{YN}.
For the example we have introduced,
since $H^{(1)}\propto\tanh(p\xi+q\eta)$,
the perturbation term will be proportional to
${\rm sech}^2(p\xi+q\eta)$.
This shows that the perturbation term results exponential
decay or amplification of the pulses monotonously.

\section{Results of Numerical Analyses}
In this section,
we shall first present the localized solution of (\ref{DS1-ALL})
used as initial-boundary conditions,
and then show the results of the numerical
analyses of the perturbed DS1 equation (\ref{all_DSP0}).
As we have pointed out in the Ref.\ \CITE{NY2},
it is inevitable to analyze the
DS1 equation numerically in the case of dromion solutions,
because of inexistence of the Hamiltonian and complexity of
them.
\subsection{Localized structures as an initial-boundary-condition
and the numerical method}
Since we are going to focus on the stability of
a dromion against perturbations,
hereafter we restrict ourselves to the initial condition
in which only one localized pulse appears.
First, we shall show the explicit form of the one dromion
solution for the DS equations (\ref{DS1-ALL}).
Setting the unknown variables as\cite{HH}
\begin{mathletters}
\label{ONEDORO-DEFS}
\begin{equation}
A=G/F,\qquad U=2(\log F)_{xx}, \qquad V=2(\log F)_{yy},
\label{ONEDORO-DEPVS}
\end{equation}
where the functions $F$ and $G$ are
\begin{equation}
\left.\matrix{
F&=&1+\exp (\eta_1+\eta_1^*)+\exp (\eta_2+\eta_2^*)
    +\gamma\exp (\eta_1+\eta_1^*+\eta_2+\eta_2^*),\cr
G&=&\rho\exp (\eta_1+\eta_2).\hfill\cr
}\right.
\label{ONEDORO-NEWVS}
\end{equation}
Writing $x=x_1$ and $y=x_2$,
we have defined the quantities $\eta_1$ and $\eta_2$ as
\begin{equation}
\eta_j=\Big[k_r^{(j)}+ik_i^{(j)}\Big]x_j
        +\Big[\omega_r^{(j)}+i\omega_i^{(j)}\Big]t,
\quad(j=1,2).
\label{ONEDORO-EXPS}
\end{equation}
\end{mathletters}%
{}From (\ref{DS1-ALL}) and (\ref{ONEDORO-DEFS}),
we have the following relations between parameters
\begin{mathletters}
\label{ONEDORO-PARAS}
\begin{equation}
\left.\matrix{
|\rho|=2\sqrt{2k_r^{(1)}k_r^{(2)}(\gamma-1)}, \hfill\cr
\omega_r^{(j)}=-2k_r^{(j)}k_i^{(j)}, \qquad
\omega_i^{(1)}+\omega_i^{(2)}
          =\sum_{j=1,2}\Big[k_r^{(j)2}-k_i^{(j)2}\Big].
\hfill\cr}\right.
\end{equation}
The boundary conditions for $U$ and $V$ are given by
\begin{equation}
U\Big|_{y=-\infty}={8k_r^{(1)2}\exp(\eta_1+\eta_1^*)\over
                  [1+\exp(\eta_1+\eta_1^*)]^2},
\quad
V\Big|_{x=-\infty}={8k_r^{(2)2}\exp(\eta_2+\eta_2^*) \over
                      [1+\exp(\eta_2+\eta_2^*)]^2}.
\label{ONEDORO_POTS}
\end{equation}
\end{mathletters}%
{}From (\ref{ONEDORO-NEWVS}) and (\ref{ONEDORO-EXPS}),
we can find that there are five free parameters in this solution.
The quantity $\gamma$  determines the amplitude of the
single pulse of $|A|$;
The parameters $k_r^{(j)}$ is the width of the pulse of
$|A|$ in the $x$ direction (when $j=1$) and $y$
direction ($j=2$),
while $k_i^{(j)}$ give the velocities in each of the directions.
Hereafter in this section, we are going to use the initial and
boundary conditions given by these (\ref{ONEDORO-DEFS}) and
(\ref{ONEDORO-PARAS}).
\par
Next we rewrite the equation to be analyzed as:
\begin{equation}
\left\{\matrix{
i A_t+A_{xx}+A_{yy}+(U+V)A-i\tilde H(x,y)A=0,\cr
U_{y}=(|A|^2)_{x},\quad V_{x}=(|A|^2)_{y}, \hfil\cr}\right.
\label{EQS_TOBE_SIMD}
\end{equation}
which is derived from (\ref{all_DSP0})
by introducing the new variables
$U$ and $V$ by
$$
U=Q_x-|A|^2,\quad V=Q_y-|A|^2.
$$
As we have considered in the previous section, the perturbation
term is chosen as $-i\alpha A{\rm sech}^2{(\mu_xx+\mu_yy)}$,
where $\alpha$ is the degree of perturbation,
and $\mu_x$ and $\mu_y$ determines the width and the direction
of the step on the bottom.
\par
Now we have to choose the five parameters in the one dromion
solution,
$\gamma$, $k_r^{(j)}$ and $k_i^{(j)}$ ($j=1,2$),
and the three which determines the perturbation term,
$\alpha$, $\mu_x$ and $\mu_y$.
Since we are going to consider a perturbation for an identical
dromion,
we use fixed values for $\gamma$ and $k_r^{(j)}$, ($j=1,2$)
all through the simulations.
The values are $\gamma=3$, $k_r^{(j)}=0.6$, ($j=1,2$).
We choose suitable values of $\alpha$, $k_i^{(j)}$, $\mu_x$
and $\mu_y$ for each case of the simulations.
\par
Finally, we show the numerical method briefly.
The space derivatives in (\ref{EQS_TOBE_SIMD}) is performed by using
the psudospectral method \cite{TA} with periodic
boundary condition.
Integration of the first equation of (\ref{EQS_TOBE_SIMD})
in terms of $t$ is performed by
the Burilsh and Store method \cite{Press}
and the fourth order Runge-Kutta method
with appropriate accuracy of adaptive step size control.
The second equation of (\ref{EQS_TOBE_SIMD}) is calculated by
the fourth order Runge-Kutta method with boundary conditions
which are presented in (\ref{ONEDORO_POTS}).
When the midpoint value between meshes is necessary,
we use cubic spline.
The detailed discussion of the scheme of the calculation is
described in the ref.\ \CITE{NY2}.
\par
\subsection{The case when the dromion traverse the step}
The first analysis is for the case of
$k_i^{(1)}=k_i^{(2)}=0.5$, $\mu_x=-\mu_y=0.4$.
In this case, the contours along the step on the bottom
runs parallel to the line $x=y$,
which is chosen as the trail of the single dromion in the
initial state travels along.
This means that the dromion undergo the effect of perturbation
constantly.
The discussion in the previous section tells us that
the height of the peak of the variable $|A|$ is considered to
decay exponentially.
We studied the time development of the height of the peak
of $|A|$ for the various values of $\alpha$,
$\alpha=0.01$, $0.05$, $0.1$, and $0.4$.
The results of this simulation is presented in
Fig.\ \ref{FIG-TRAV}.
The data of the case $\alpha=0.4$ deviates from a straight
line with the lapse of time.
This is because the pulse decays sufficiently
(less than 1\% in height),
and hence it cannot be distinguished from the ripples
distributed other places on the plane.
We can immediately see that the result is consistent with the
discussion above,
and can also corroborate the accuracy of the scheme.
\par
\subsection{The case when dromion goes across the step}
In this subsection, we set $k_i^{(1)}=k_i^{(2)}=0.5$,
$\mu_x=\mu_y=0.4$.
This means that the initial dromion travels across the step.
The dromion feels the effect of perturbation
only when it passes near the origin.
If dromions are not stable structure, they are supposed to break
into small fractions due to the effect of perturbations,
even after the effect vanishes.
On the other hand, if they are stable,
the opposite will occur;
A localized pulse travels stably
after disappearance of the effect of perturbation,
though the height of its peak changes by the effect.
\par
The numerical analyses of the time development of $|A|$ is
performed both for positive and negative values of $\alpha$.
In Fig.\ \ref{FIG-ACR-DEC} and \ref{FIG-ACR-AMP},
the time developments of the height of the peak is shown.
The height of the dromion changes near the origin,
only where the perturbation effect appears,
but it keeps the same value after the dromion passed there.
We displayed in Fig.\ \ref{FIG-SURF} the surface pictures
of the main flow before and after the pulse get through
the perturbation area.
The dromion keeps a feature as an localized pulse,
though it experienced perturbations.
\par
Next we select various values of the velocity of dromion
and investigate how the effect of perturbation affects on
the pulse.
We consider a relation between velocities of dromion
and decay ratio, the height of the peak of $|A|$ after dromion
go through the perturbation normalized by its initial value.
We choose a fixed value of $\alpha=0.1$, and different values of
velocities $k_i^{(1)}=k_i^{(2)}=N/16$, $N=8,\ldots ,12$.
The relations between logarithm of the decay ratio and the
inverse of the parameter $k^{(j)}_i$, ($j=1,2$) is presented in
Fig.\ \ref{FIG-VELO}.
This shows that there is a linear relation between them.
\par
This result is explained as follows.
As we have discussed in the previous section,
a wave with small amplitude decays as
$F\sim\exp(-\alpha J(x,y)t)$.
In this case, $F(t)$ is the maximum value of $|A|$ and
$J(x,y)={\rm sech}^2[\mu(x+y)]$.
Considering the dromion traveling along the line $y=x$
with velocity $v$,
we have
$dF|_{x\to x+dx}
 \cong -\alpha Fv^{-1}{\rm sech}^2(2\mu x)\,dx$.
Integration of this equation gives the function $F$ to be
$F\sim\exp[-\alpha(2\mu v)^{-1}\int d\xi{\rm sech}^2\xi]$.
Then the decay ratio, $D\equiv F(\infty)/F(-\infty)$ satisfies
a relation $\ln D\sim -(\alpha/\mu)v^{-1}$,
and this shows a linear relation between $\ln D$ and $1/v$.

\section{Concluding Remarks}
We have investigated time developments of surface waves
described by the DS1 equations with perturbation numerically.
The main results of the calculations tell that
dromions decay or are amplified when they feel perturbation,
and they travel stably after the effects of perturbations
disappear.
In addition, if the time during which the dromions
undergo the effects of perturbation becomes longer,
the degree of decay will be larger.
Taking these numerical results into account,
we can conclude that dromions are quite stable structures in
two-dimensional space.
\par
Now we would like to consider the process
under which the dromions travel stably in spite of
the existence of perturbations.
As we have discussed in the previous paper,
the mean flow $U$ and $V$ play the important roles
in (\ref{DS1-ALL}) or (\ref{EQS_TOBE_SIMD}).
They behave as attractive potentials of the Schr\"odinger
equation,
and hence preserve the localized structures of the main flow.
Because the cross point of the peaks of $U$ and $V$ is
the most attractive point in the system\cite{NY2},
and it gathers structures of the main flow.
The stability of the dromion solutions owes much
to the mean flow.
\par
We can say that, as long as
attractive points generated from the mean flow
$U$ and $V$ exists,
localized structures of the main flow will keep on existing.
The stability revealed in this paper will assure
observations of dromions in existing system,
such as fluid dynamics and plasma systems,
and this will be a future problem.

\section*{Acknowledgments}
The authors would like to express their sincere gratitude to
Dr.\ Takeshi Iizuka for valuable comments
on nonlinear phenomena in inhomogeneous systems.
They are grateful to Professor Junkichi Satsuma and
for continual encouragements.
Thanks are also due to Professor Miki Wadati
for use of his computers.

\newpage
\begin{figure}
\caption{
A semi-log graph of time development of the highest part of
$|A|$ of one dromion for the perturbed DS equations
when the dromion traverse the step.
The height is normalized by that of the solution with no
perturbation.
The values of $\alpha$ are $0.01$, $0.05$, $0.1$ and $0.4$.
The peak decays exponentially.
The reason of the deviation from a straight line in the case
of $\alpha=0.4$ is that the pulse gets as low as ripples
distributed other places on the plane.}
\label{FIG-TRAV}
\end{figure}
\begin{figure}
\caption{
The developments of the highest part of $|A|$, when the dromion
goes across the step.
The values of $\alpha$ are $0.01$, $0.05$, $0.1$ and $0.5$.
Since $\alpha$ is negative, the dromion goes into the direction
where the depth of the fluid grows deeper, and undergo decay.
The modification in height occurs when the dromion passes
near the origin ($t=0$), where the perturbation term has finite
value.}
\label{FIG-ACR-DEC}
\end{figure}
\begin{figure}
\caption{Time developments of highest part of $|A|$.
The motion of the dromion is the same as those presented in
Fig.\ \protect\ref{FIG-ACR-DEC}.
The values of $\alpha$ are $-0.01$, $-0.03$ and $-0.05$.
and this means that the dromion climbs up the step.
The dromion is amplified, and the change of height occurs
around the region where the perturbation exists as in
Fig.\ \protect\ref{FIG-ACR-DEC}.}
\label{FIG-ACR-AMP}
\end{figure}
\begin{figure}
\caption{
Typical surface figures of the envelopes of the main flow,
when the dromion goes across the step and decays.
The parameter is $\alpha=0.2$, and the time is
(a) $t=-5.5$, (b) $t=-1.0$ and (c) $t=4.5$.
Though the height of the peak is lowered, the pulse keeps the
characteristic as a localized structure after the effect of
perturbation.}
\label{FIG-SURF}
\end{figure}
\begin{figure}
\caption{The relation between velocity of dromion and decay ratio.
The vertical axis is the logarithm of the decay ratio and
the horizontal one inverse of $k^{(j)}_i$, ($j=1,2$).
The parameter is $\alpha=0.1$.
The straight line is drawn by the method of least square.
The dromion which travels faster undergo less decay.}
\label{FIG-VELO}
\end{figure}
\newpage
\begin{table}
\caption{
Main results of the expansion amplitudes in
(\protect\ref{EXPAND_all})
from the reductive perturbation method.
The variables $A_j$ ($j=0,1$) and $B_j$ ($j=0,1,2$) are
functions of $\xi$, $\eta$ and $\tau$,
and $Z$ is defined by $Z\equiv z+H^{(0)}$.
The quantity $v$ is the group velocity
$\partial\omega/\partial k$ and
$c$ is the phase velocity $\omega/k$.
The values with negative $\ell$ can be derived from the
relations $\phi^{(n)}_{-j}=\phi^{(n)\ast}_{j}$ and
$\zeta^{(n)}_{-\ell}=\zeta^{(n)\ast}_{\ell}$.}
\label{TAB-RES-RPM}
\begin{tabular}{cll}
$(n,l)$ & $\phi^{(n)}_\ell(\xi,\eta,z,\tau)$
        & $\zeta^{(n)}_\ell(\xi,\eta,\tau)$ \\
\tableline
$(1,0)$ & $A_0(\xi,\eta,\tau)$ & $0$ \\
$(1,1)$ & ${\displaystyle\strut
           A_1(\xi,\eta,\tau)\cosh(kZ)
           \over\displaystyle \cosh(kH^{(0)})}$
        & ${\displaystyle\strut i\omega A_1
           \over\displaystyle g(1+K)}$\\
$(2,0)$ & $B_0(\xi,\eta,\tau)$
        & ${\displaystyle\strut vA_{0\xi} \over\displaystyle g}$ \\
$(2,1)$ & ${\displaystyle\strut
           B_1(\xi,\eta,\tau)\cosh(kZ)
           -iA_{1\xi}Z\sinh(kZ)
           \over\displaystyle \cosh(kH^{(0)})}$
         & ${\displaystyle\strut i\omega B_1
           +A_{1\xi}
           \Big(\sigma\omega H^{(0)}+v-
           {\displaystyle 2cK
            \over\displaystyle 1+K}\Big)
           \over\displaystyle g(1+K)}$ \\
$(2,2)$ & ${\displaystyle\strut
           B_2(\xi,\eta,\tau)\cosh(2kZ)
           \over\displaystyle \cosh(2kH^{(0)})}$
        & $-{\displaystyle\strut k^2A_1^2
            \over\displaystyle 2g(1+K)}
          +i{\displaystyle\strut 2k\sigma B_2^2
            \over\displaystyle \omega(1+\sigma^2)}$ \\
\end{tabular}
\end{table}

\end{document}